# High speed coherent one-way quantum key distribution prototype

Damien Stucki[1], Claudio Barreiro[1], Sylvain Fasel[1], Jean-Daniel Gautier[1], Olivier Gay[2], Nicolas Gisin[1], Rob Thew[1], Yann Thoma[1], Patrick Trinkler[2], Fabien Vannel[1], Hugo Zbinden[1]*

[1] Group of Applied Physics, University of Geneva, 1211, Geneva 4, Switzerland
[2] id Quantique SA, Chemin de la Marbrerie 3, 1227, Carouge, Geneva, Switzerland
* e-mail: Hugo.Zbinden@unige.ch

**Abstract**

We report the first real world implementation of a Quantum Key Distribution (QKD) system over a 43dB-loss transmission line in the Swisscom fibre optic network. The QKD system is capable of continuous and autonomous operation and uses the coherent one-way (COW) protocol. This system brings together three key concepts for future QKD systems: a simple high-speed protocol; high performance detection; and integration, both at the component level as well for connectivity with standard fibre networks. Here, we show laboratory and field trial results for this system. The full prototype version uses InGaAs/InP avalanche photodiodes (APDs) and was laboratory tested up to 150km, with a 10-hour exchange averaging around 2kbps of real-time distilled secret bits over 100km. In the field trials, we obtained average distribution rates, during 3 hours, of 2.5bps over a 43dB-loss line of 150km, when using superconducting single photon detectors (SSPDs).

**Introduction**

Quantum Key Distribution (QKD) provides a means of generating secret random bit strings, keys, at two distant locations, such that their secrecy is guaranteed by the laws of quantum physics[1]. Invented in 1984, QKD witnessed a rapid development during the 1990s culminating in the first commercial prototypes a few years ago[2]. Since then efforts have focused on three crucial aspects: improved components; better protocols; and more generally, system integration, for continuous operation in real world networks. The first targets increasing maximum transmission range and bit rates using e.g. high performance superconducting detectors[3]. The second has resulted in the development of new schemes that render faint laser systems less vulnerable against photon number splitting attacks[4,5,6,7,8]. Finally, the continuous operation, standardisation and the development of secure networks and network protocols[9,10] will be necessary for eventual commercial success. The goal of our work has been to combine these three lines of research.

The performance of a QKD system can be described by the secret bit rate $K = Rr$, where $R$ is the sifted bit rate and $r$ denotes the secret key fraction[11]. In principle, $R$ is proportional to the pulse rate of the source $v_s$ and the average number of photons per pulse $\mu$. Increasing the rate (or $\mu$) is a straightforward way to increase $R$. Pulse rates of more than 1GHz[12,13,14] and even up to 10GHz, though only over a fraction of a second[3], have already been demonstrated by some groups. However, $r$ depends on the potential information of an eavesdropper and in the case of faint laser systems, which are vulnerable to photon number splitting (PNS) attacks, depends on $\mu$. The implementation of faint laser schemes, like "Differential Phase Shift" (DPS)[4] or SARG[5], or decoy states[15], which are resistant against PNS attacks, allows one to increase $\mu$ and therefore, the maximum distance and secret bit rate. Finally, the quantum bit error rate (QBER) is dependent on the ratio between the detector noise and transmission loss. Therefore the detector noise limits the maximum transmission distance, hence the interest in low noise superconducting detectors[3,16].

We have developed a QKD prototype based on a PNS-attack resistant protocol called *coherent one-way*[7] (COW) that takes advantage of high pulse rates, continuous and automated operation, and free-running, low noise, detectors. We report on laboratory tests up to 31dB (>150km of standard fibre), with a 10-hour exchange averaging around 2kbps over 21dB (100km) using InGaAs/InP avalanche photodiodes (APDs). We also performed a field trial, over the Swisscom fibre optic network between Geneva and Neuchâtel, a physical distance of 110km with a fibre transmission distance of 150km in a high loss (43dB) line, corresponding to over 200km of standard fibre. Using superconducting single photon detectors (SSPDs), we found average distribution rates of 2.5bps with a fully autonomous QKD system.

**The Coherent One-Way QKD Protocol**

In the COW QKD protocol, logical bits are encoded in time[7]. A sequence of weak coherent pulses is tailored from a CW-laser with an external intensity modulator (see Figure 1). The emitter, Alice, encodes using time slots (separated by T) containing either 0-pulses, no light, or µ-pulses, with a mean number of photons of µ<1. The logical bit $0_L$ ($1_L$) corresponds to a sequence 0-µ (µ-0). For security reason, we also send µ-µ decoy sequences. The receiver, Bob, registers the time-of-arrival of the photons on detectors $D_B$ for the data line and $D_M$ for the monitoring line. The $D_B$ times give the raw key from which Alice and Bob extract the net key. The security is guaranteed by checking for detections on $D_M$, for decoy sequences and logical sequences $1_L 0_L$, at random times using an unbalanced interferometer that has a path-length difference of T. The interferometer is only used to estimate the information of the eavesdropper and cannot introduce errors on the key.

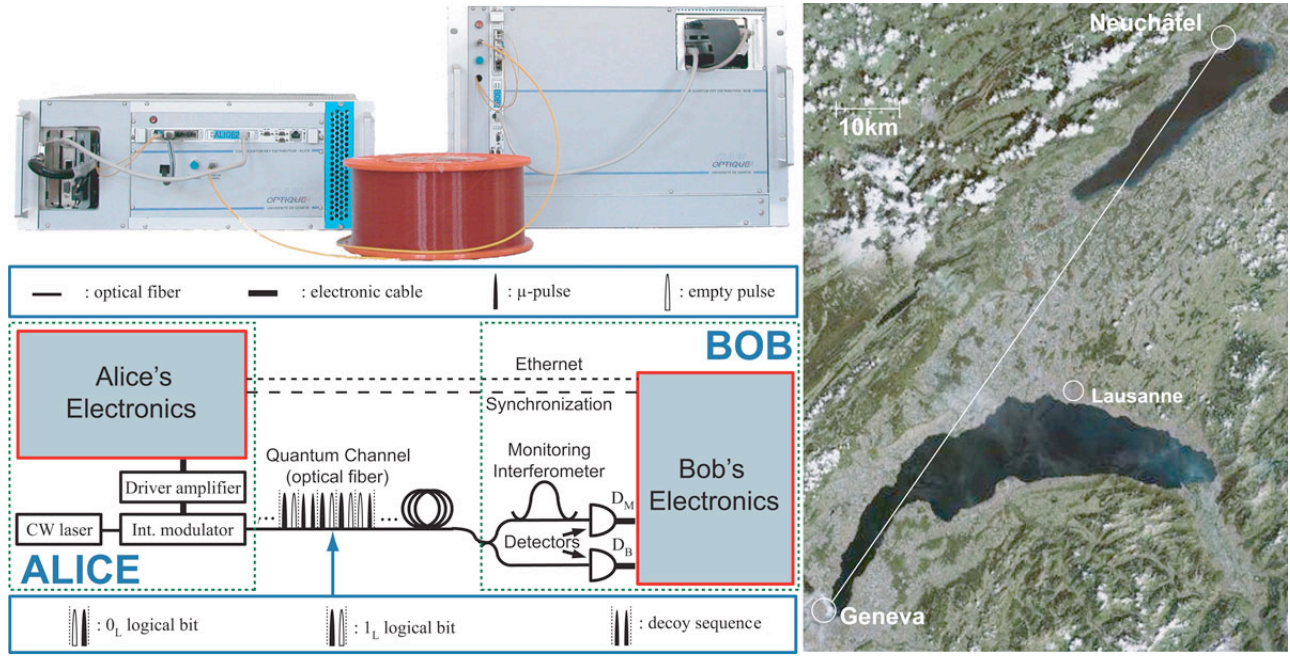

**Figure 1 Coherent one-way protocol.** A schematic of the setup is shown (bottom left) with a photograph (above) of the final 19" rack-mountable system. On the right we see a satellite image showing the two cities used for this QKD field trial.

**Security**

Proving the security of the COW protocol remains a work in progress. The standard methods for proving the security of QKD protocols were, so far, developed for protocols in which the quantum symbols are sent one-by-one (e.g., qubits in the BB84[17], B92[18], SARG[5] protocols). The COW protocol, however, does not use this symbol-per-symbol type of coding and the standard security proofs do not apply in any straightforward way. To the contrary, the COW protocol is a so-called *distributed-phase-reference* protocol, like DPS, which relies on the coherence between successive non-empty pulses to ensure the security of the protocol. So far, the security of the COW protocol has been proven against the Beam-Splitting Attack (BSA)[7,8,19,20] and some intercept-resend attacks[7,19] and against arbitrary collective attacks, under the assumption that Bob receives at most one photon per pulse. Here we use an estimate on Eve's information:

$$I_{AE}(\mu) = \mu(1-t) + (1-V)\frac{1+e^{-\mu t}}{2e^{-\mu t}},$$

where the first term corresponds to BSAs and the second to intercept-resend attacks. $t$ is the line transmission and $V$, the interferometer's visibility. From this we can also determine an upper bound on the secret key rate:

$$K = R(\mu, \nu_s)\left[1 - h(Q(\mu)) - I_{AE}(\mu)\right].$$

The second term corresponds to the secret key fraction $r$. $h(Q(\mu))$ is the Shannon entropy, for a given QBER, that is related to the minimum fraction of bits lost due to error correction (EC).

**Description of the prototype**

The system has been developed to be rack mountable in two 19 inches boxes. Two fibres link them together; the first used for the quantum channel, the second for the classical communication, synchronisation and pre-sifting, which we explain momentarily. Each box is connected to the Internet for the classical processing of the raw key. Alice and Bob have similar structures: optical and electronics modules with Field Programmable Gate Arrays (FPGAs) and embedded computers.

The optical modules are very simple (see Figure 1). Alice's optics consists of a CW DFB telecom laser diode at 1550nm (Thorlabs, 1554.94-20) and a Lithium niobate intensity modulator (Avanex, 10Gbits/s) to tailor the sequence of pulses. After the intensity modulator a 50/50 coupler splits the beam in two where a PIN InGaAs detector monitors the power so as to set the variable attenuator (OZ Optics, DD-600-xx) to the appropriate mean number $\mu = 0.5$ photons per pulse, independent of fibre length. On Bob's side, a 90/10 coupler sends 90% of the photons directly to a single photon detector $D_B$ (bit or data channel). The remaining 10% of the photons go through an unbalanced Michelson interferometer and are detected by $D_M$ (monitor channel). The interferometer is passively temperature stabilised.

Alice and Bob's units use embedded computers to monitor and control the full system. Each computer has several tasks: one process supervises all the subsystems and the communication between Alice and Bob; and a second, controls various regulation tasks, e.g. for Alice, the current of the DFB laser. Another processor is used to handle communication between the embedded computer and FPGA. The optical system is driven at a rate of 625MHz with a subsequent logical bit rate of 312.5Mb/s. The outgoing signal of the FPGA is shaped via a homemade ECL electronic circuit and driver amplifier (Picosecond, 5865, 12.5Gb/s) producing 300ps optical pulse widths. A synchronisation and pre-sifting (from Bob to Alice) signals are combined using a WDM and sent through the second (classical) optical channel.

A Quantum Random Number Generator (QRNG) (id Quantique, Quantis OEM module) produces random bits at a rate of 4Mbits/s. However, logical bits, at a rate of 312.5Mb/s have to be generated. We use the output of this QRNG to frequently seed a pseudo-random number generator (PRNG) implemented inside the FPGA. This PRNG is made of a 32-bit Linear Feedback Shift Register (LFSR) and produces pseudo random sequences of ($2^{32}$ - 1) bits, a period of around 13s. In our system, the QRNG supplies a new seed to the PRNG every 64μs thus avoiding sequence repetitions. This solution allows us to obtain the 312.5Mb/s rate demanded by our system. Decoy sequences are sent whenever the 1010 string is found at the output of the random generator. The generation of quantum random bits at GHz rates is currently an open problem. *n.b.* The COW protocol only requires a RNG on Alice's side.

On Bob's side, InGaAs APDs are Peltier cooled to around -50°C and provide a simple, reliable and cheap detection system. Normally, these detectors are used in a so-called gated mode but in our case we use them in the recently developed free running mode[21]. We obtain dark count rates and quantum efficiencies of the order of $10^{-6}$ per ns and 10%, respectively. To limit the after-pulse probabilities to below $10^{-5}$ per ns, dead times of the order of 30μs have been applied that limits the maximal detection rate to ~30kHz, which is not a problem for longer distances, but a limiting factor at short distances. At short distances up-conversion based detection schemes, with detection rates up to 10MHz[12], could provide an interesting alternative. For the long distance experiments we use SSPDs, because of their low noise capabilities[22]. The sub-4K operating temperature is certainly a handicap for a commercial system, however, the system is a fibre coupled SSPD in a cryogen free cryostat, developed within the European project Sinphonia[23], which makes their application rather straightforward in a laboratory environment. These detectors have quantum efficiencies up to 2.5% with noise levels less than 10 counts per second that can be adjusted via a bias current. Currently though, their dependence on polarisation, which can reduce the quantum efficiency by 50%, is problematic for fibre transmission.

**Prototype operation**

Before the operation of any QKD system an installation procedure must be undertaken that includes, checking component operation, such as the efficiency of detectors and the optimum parameters for the intensity modulator as well as network characteristics like the total attenuation of the fibre transmission line. However, from this point, all the initialisation and synchronisation for the exchange is performed automatically by the COW system. In particular, the following steps are carried out:

*a) Identification of bit numbers:* Alice sends a series of different pulse patterns over the synchronisation channel, during initialisation, so that Bob can identify the pulse numbers, i.e. the length difference between the classical and quantum fibres.

*b) A fine temporal tuning of the detection window.* The delay, between synchronisation and the detector output pulses, is scanned to maximise the count rates, which also reduces noise and crosstalk between adjacent pulses.

*c) Wavelength adjustment and interference visibility.* Random pulse sequences are sent over the quantum channel, the wavelength of the laser is scanned and the count-rate on monitoring detector recorded. The resulting interference fringe is used to calculate the visibility and set the wavelength for destructive interference.

Figure 2 shows an example of the data on the monitoring line. On the left hand side we see the noise rate and then the rate during a scan of the laser wavelength when random pulse sequences are sent. The maximum and minimum levels are then determined and if the minimum level is sufficiently close to the noise level, say a visibility >97%, then the key exchange is started. The laser wavelength is continuously tuned in order to minimise the monitoring count rate and keep it above 95%. If for some reason the visibility drops below this limit, a complete scan is automatically performed again and the key exchange restarts.

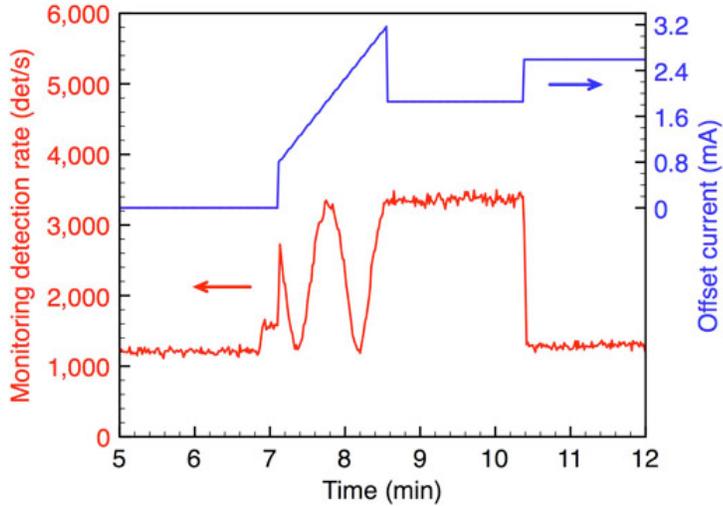

**Figure 2 Monitoring line.** Alice's laser wavelength (current) is periodically scanned to optimise the visibility of Bob's monitoring interferometer.

To perform the key exchange Alice randomly sends bits encoding $0_L$, $1_L$ and decoy sequences. Bob registers all detection times for the data and the monitoring detector. He then announces for which bit there was a detection on $D_B$, as well as the times when he got clicks on the monitoring detector, $D_M$. In order to save memory space, Alice uses this *pre-sifting* information to immediately delete the bit values that are not needed. She then checks the security using the relevant monitoring detections, i.e. detections that correspond to interfering pulses. If the associated visibility is over 95%, the key exchange continues. Alice then tells Bob, which detections, corresponding to decoy sequences, have to be removed from his data. Thus, Alice and Bob finally have a shared stream of bits, the sifted key.

The distillation software then corrects the errors in the sifted key and applies the privacy amplification algorithm. The error correction is based on the Cascade algorithm[24]. As the bit exchange is continuous, the distillation software has to run in parallel, in real time, and treats blocks of data, $2^{13}$ or $2^{14}$ bits, depending on the distance, so as to ensure a good efficiency. Eve's information is removed through privacy amplification implemented using hashing functions based on Toeplitz matrices[25]. Finally, all information exchanges over the classical channel during the distillation procedure are securely authenticated using a Wegman-Carter type scheme, implementing universal hashing functions[26].

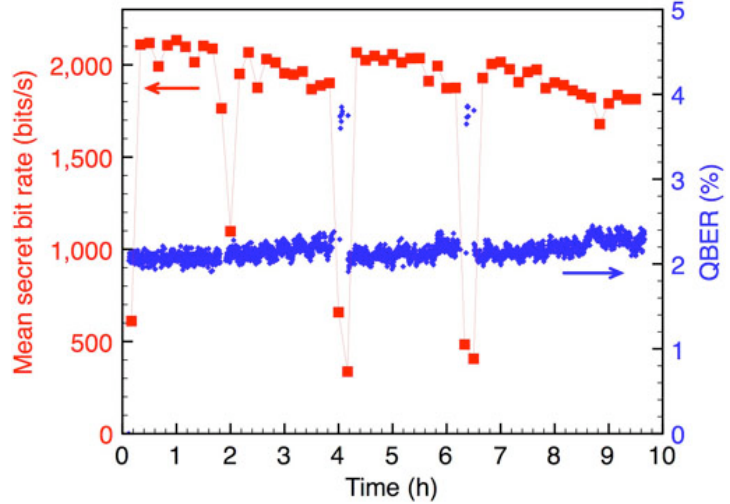

**Figure 3 Secret bit rate and QBER vs Time.** A 10-hour key exchange with 25km of fibre plus 15dB attenuation, equivalent to around 100km of standard fibre. We show the rate averaged over 10 minute intervals. The drops in rate denote periods of auto re-alignment when key generation is interrupted.

## Results

We have performed a series of measurement in the lab using InGaAs detectors. We introduced 25km of fibres and a variable attenuator between Alice and Bob and we started all key exchanges with the complete initialisation and synchronisation procedure. In Figure 3 we can see the results of a key exchange with 21dB loss, corresponding to >100km of standard fibre. We find a bit rate mean secret bit rate of around 2kbps, after the distillation and authentication process. This figure illustrates the system's capability for continuous operation and its ability to automatically recover when the QBER increases excessively.

In Figure 4 we present the results obtained with 25km of fibre (around 6dB) with additional loss from an attenuator. Each point of the graph gives the mean secret bit rate and the QBER obtained over one hour. The rate is the direct output from the distillation software, it is not an evaluation based on the raw detection rate, the QBER and the visibility of the interferometer. For short distances and low losses, the rate is limited by detector saturation, leading to the almost constant rate up to 21dB. Nonetheless, we are able to distribute > 50bps over a 31dB-loss line. Note, at 6dB we see a sharp increase in the QBER that is due to charge persistence in the detectors. Due to the high photon flux, there is an increased probability that a photon arriving before the detector is "on" provokes a detection avalanche. This leads to additional errors and an increase in QBER.

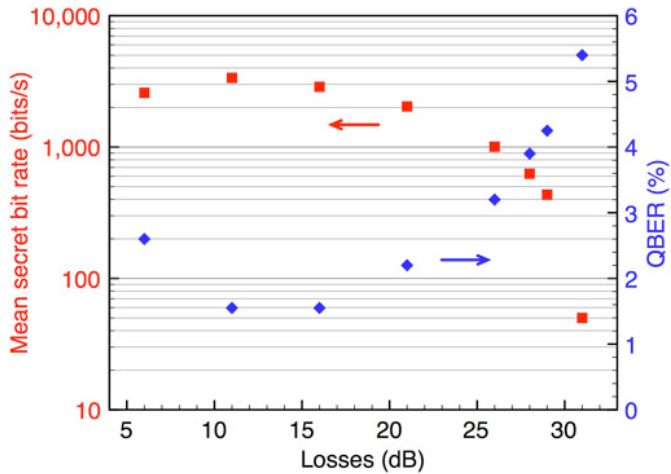

**Figure 4 Secret bit rate and QBER vs Distance.** Rates and errors as a function of the loss, or distance. 31dB corresponds to ~150km of standard fibre.

## Field Measurements

We have also performed field measurements using a fibre link between our lab in Geneva and a lab of our colleagues at the University of Neuchâtel using the Swisscom fibre network. While Alice travelled to Neuchâtel, we connected Bob to SSPD detectors situated in the Geneva lab. Neuchâtel is some 110km beeline from Geneva, the fibre, however, had a length of about 150km and rather high loss of 43dB, due to many connections in between.

In Figure 5a we show the data for an exchange over 3.5 hours, featuring an average secret bit rate of roughly 2.5 bits per second with a reasonable QBER of about 5%. At this long distance, the count rates on the monitoring line become very small (< 1 per 10s) and the statistics aren't sufficient to correctly establish the value of the visibility and continuously adjust the laser wavelength. Therefore, while this is probably the longest key exchange ever, its security is questionable.

We finally used the same installed 150km fibre and mimicked key exchanges over intermediate distances by increasing the pulse energy of the outgoing pulses at Alice. The results are summarised in Figure 5b, where you can see the secret key rate as a function of fibre loss. We see some fluctuations about the theoretical value that is in part due to detector efficiency's sensitivity to polarisation. At shorter distances < 20dB the rate is again saturated, although here, it is due to the EC and the classical communication time between Alice and Bob who are still separated by 150km of fibre in this case. It turns out, that for longer distances, lower detector efficiencies, and hence lower noise, can lead to improved overall rates. To see this, we lowered the bias current on the detectors and found the resulting (circled) distribution rates of 2.5bps at 42-43dB, i.e. over 200km of standard fibre loss.

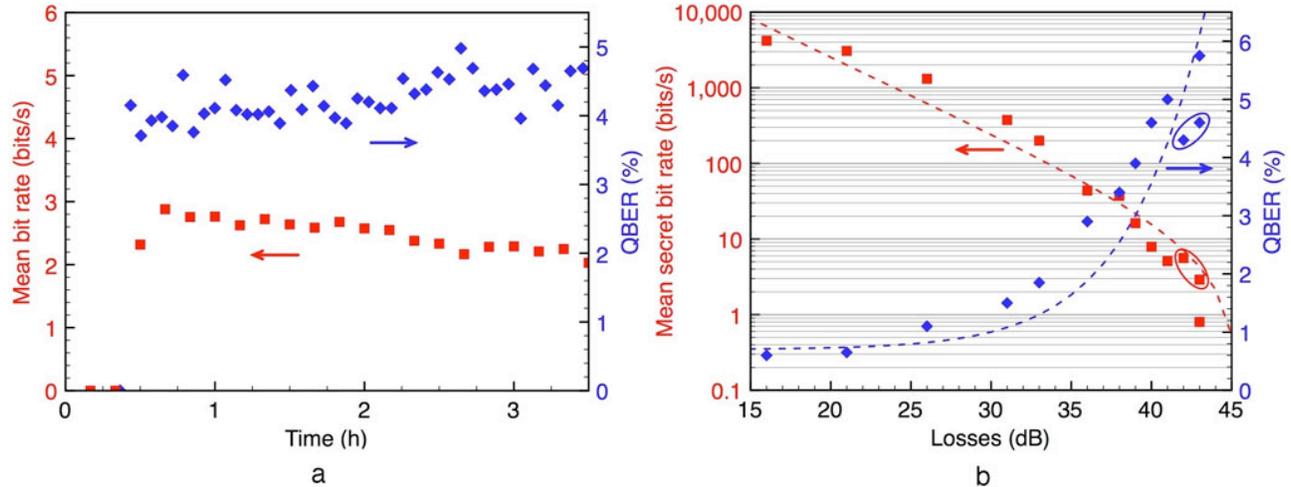

**Figure 5 Long distance QKD.** a) Continuous operation over 3.5 hours for a 43dB-loss line. b) Rates and errors as a function of loss in the installed, Geneva-Neuchâtel, fibre.

**Conclusion**

We have presented results for a prototype QKD system in a real world implementation of QKD over distances corresponding to 43dB in the Swisscom fibre optic network between the cities of Geneva and Neuchâtel. The QKD system has integrated optics, electronics and software that allows for continuous and autonomous operation. We have used the coherent one-way (COW) protocol, which is a protocol that was invented with this specific goal in mind. Here, we have shown laboratory and field trial results for this system. The full prototype version uses InGaAs/InP avalanche photodiodes (APDs) and was laboratory tested up to 31dB (~150km), with a 10-hour exchange averaging secret bit rates over 2kbps for 21dB. In the field trials, with superconducting single photon detectors, we find average bit rates of 2.5bps for 43dB. The results presented here are a major step forward towards inter-city QKD for future quantum networks.

**Acknowledgements**

The authors thank: C. Branciard & V. Scarani for useful discussions concerning QKD security; D. Salart & N. Walenta for their assistance with the field installation and the SSPDs; Y. Hasani, T. Lörunser, from ARC GmbH, with the RNG; CES SA with the FPGA cards; M. Rufer & D. Hofstetter for the use of their lab at the University of Neuchâtel; and Swisscom for access to their fibre link. Financial support is acknowledged from the EU projects, SECOQC and SINPHONIA as well as the Swiss NCCR "Quantum Photonics".